\documentclass[lettersize,journal]{IEEEtran}
\usepackage{amsmath,amsfonts}
\usepackage{algorithmic}
\usepackage{algorithm}
\usepackage{array}
\usepackage[caption=false,font=normalsize,labelfont=sf,textfont=sf]{subfig}
\usepackage{graphicx}
\usepackage{float}
\usepackage{subfig}
\usepackage{textcomp}
\usepackage{stfloats}
\usepackage{url}
\usepackage{verbatim}
\usepackage{graphicx}
\usepackage{cite}
\begin{document}

\title{Deep Learning for CSI Feedback: One-Sided Model and Joint Multi-Module Learning Perspectives}

\author{Yiran Guo, Wei Chen,~\IEEEmembership{Senior Member,~IEEE,} Feifei Sun, Jiaming Cheng, \\Michail Matthaiou,~\IEEEmembership{Fellow,~IEEE}, and Bo Ai,~\IEEEmembership{Fellow,~IEEE}

\thanks{Yiran Guo, Wei Chen, Jiaming Cheng and Bo Ai are with the School of Electronic and Information Engineering, Beijing Jiaotong University, China.

Feifei Sun is with the Samsung R\&D Institute China-Beijing, China.

Michail Matthaiou is with the Centre for Wireless Innovation (CWI), Queen’s University Belfast, UK.}
}

\maketitle

\begin{abstract}
The use of deep learning (DL) for channel state information (CSI) feedback has garnered widespread attention across academia and industry. The mainstream DL architectures, e.g., CsiNet, deploy DL models on the base station (BS) side and the user equipment (UE) side, which are highly coupled and need to be trained jointly. However, two-sided DL models require collaborations between different network vendors and UE vendors, which entails considerable challenges in order to
achieve consensus, e.g., model maintenance and responsibility. Furthermore, DL-based CSI feedback design invokes DL to reduce only the CSI feedback error, whereas jointly optimizing several modules at the transceivers would provide more significant gains. This article presents DL-based CSI feedback from the perspectives of one-sided model and joint multi-module learning. We herein introduce various novel one-sided CSI feedback architectures. In particular, the recently proposed CSI-PPPNet provides a one-sided one-for-all framework, which allows a DL model to deal with arbitrary CSI compression ratios. We review different joint multi-module learning methods, where the CSI feedback module is learned jointly with other modules including channel coding, channel estimation, pilot design and precoding design. Finally, future directions and challenges for DL-based CSI feedback are discussed, from the perspectives of inherent limitations of artificial intelligence (AI) and practical deployment issues.

\end{abstract}

\section{Introduction}
Given the global success of fifth-generation (5G) technologies, anticipation is growing for the enhanced performance that sixth-generation (6G) would offer in a wide range of applications. Massive multiple-input multiple-output (MIMO) technology has the capability to utilize spatial resources by employing multiple antennas at both the transmitting and receiving end, thereby enhancing the spectral efficiency (SE) and system throughput. This plays a crucial role in delivering the potential of 5G and meeting its requirements. Its next-generation embodiment, namely ultra-massive MIMO technology, is also expected to be one of the key technologies underpinning 6G \cite{6G}.

\begin{figure}[!t]
\centering
\includegraphics[width=3.4in]{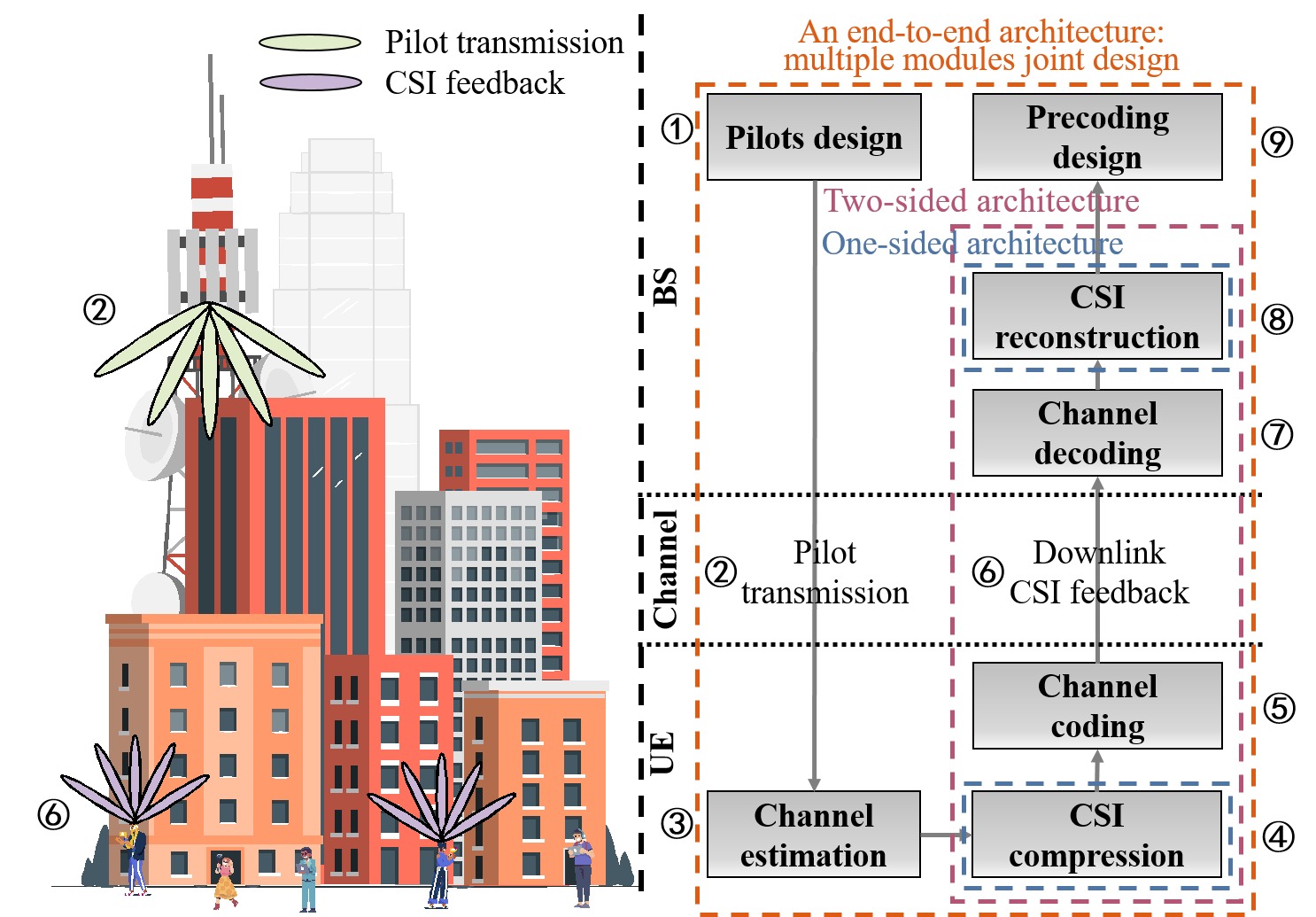}
\caption{A communication framework with CSI feedback.}
\label{scenario}
\end{figure}

To fully exploit the potential of MIMO technology, it is crucial to obtain downlink channel state information (CSI) on the base station (BS) side. In time-division duplexing (TDD) systems, the uplink and downlink channels working at the same frequency band result in channel reciprocity, which allows the downlink CSI to be obtained from the uplink CSI. However, in frequency-division duplexing (FDD) systems, the uplink and downlink channels work at different frequencies, resulting in the loss of channel reciprocity. To obtain downlink CSI, it is essential to feed the estimated downlink CSI from the user equipment (UE) back to the BS through the uplink channel, as shown in Fig.~\ref{scenario}. The feedback of CSI leads to extra overhead, which escalates significantly with the increase in the number of antennas deployed. When the uplink channel resources are limited, the challenge of CSI feedback lies in maintaining the accuracy of the feedback while minimizing the feedback overhead.

Traditional methods for reducing the CSI feedback overhead include techniques based on codebooks and compressed sensing (CS). Codebook-based CSI feedback is currently adopted in 5G systems, e.g., the Type I codebook and the Type II codebook. The pre-designed codebook is known to both the receiver and transmitter. After estimating the downlink CSI at the UE, the quantified index of the precoding matrix is computed using the CSI and the codebook, and then it is returned. However, the performance of this method is limited by the size of the codebook. As the number of antennas grows, the codebook's size grows also significantly, resulting in increased feedback overhead and more challenging searches for codewords. CS-based CSI feedback compresses the CSI information via linear projections and reconstructs the original CSI by exploiting the sparse characteristics of the channel due to the limited local scatterers. However, it heavily relies on the assumption of channel sparsity, which is not always met in practical scenarios, while the computational complexity of the iterative reconstruction algorithm is high. In the past few years, the concurrent processing abilities of graphics processing units (GPUs) have enabled the use of artificial intelligence (AI) and deep learning (DL) methods in various fields, such as image and natural language processing, leading to significant improvements in performance. Lately, AI has been utilized in the CSI feedback space to improve the precision of CSI reconstruction; see, for instance, the AI for CSI feedback enhancement in the study item of the Third Generation Partnership Project (3GPP) Release 18 \cite{AI_CSI_usecase}.

\begin{figure}[!t]
  \centering
  \subfloat[]{
    \label{one_side}
    \includegraphics[width=3.3in]{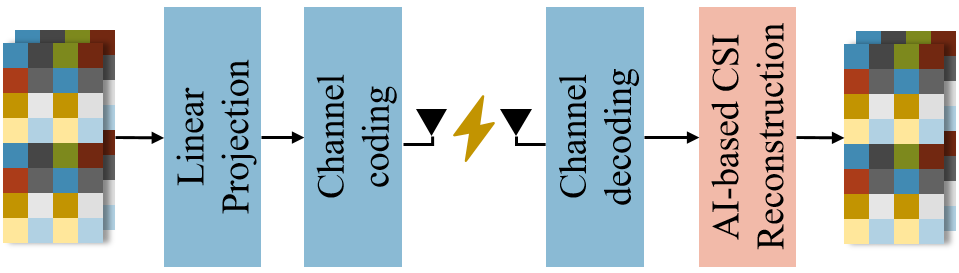}
    
  }\\
  \subfloat[]{
    \label{SSCC}
    \includegraphics[width=3.3in]{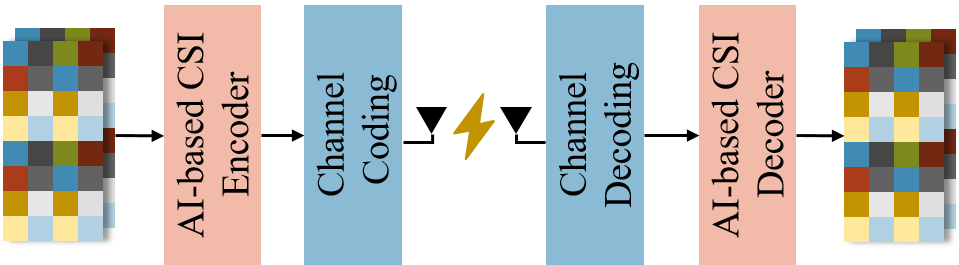}
    
  }\\
   \subfloat[]{
    \label{DJSCC}
    \includegraphics[width=3.3in]{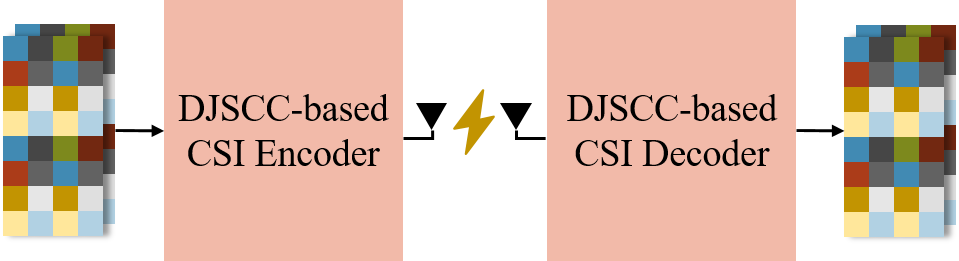}
    
  }\\
  \caption{The construction of AI-based CSI feedback: \protect\subref{one_side} one-sided AI-based CSI feedback; \protect\subref{SSCC} two-sided AI-based CSI feedback excluding channel coding module; \protect\subref{DJSCC} two-sided AI-based CSI feedback including channel coding module.}
  \label{construction}
\end{figure}

Most existing AI-based feedback methods exploit two-sided DL models, wherein the DL models on the BS side and the UE side are highly coupled, and need to be trained jointly. The two-sided DL models require collaborations between different network vendors and UE vendors in the training and inference phases, which entails various issues to be considered in order to achieve consensus, e.g., model maintenance and responsibility. The drawback of two-sided DL models can be overcome by using one-sided CSI feedback models. As shown in Fig.~\ref{construction}, one-sided CSI feedback deploys AI models only at the UE or the BS, which reduces the excessive coordination between both sides.


Apart from the CSI feedback task, AI techniques have also been applied to other modules within communication systems to tackle nonlinear and complex tasks, such as channel coding \cite{DL_channel_encoder}, channel estimation (CE) \cite{DL_CE} and precoding design. An independent design of these modules is not optimal, as the output of the previous module affects the optimization space of the subsequent modules. Therefore, the simultaneous optimization of several modules through an AI-driven end-to-end network appears more advantageous. This approach could streamline the architecture of the DL network and reduce the complexity of training by distilling task-specific semantic information from the final task.

\begin{table*}[!t]
\caption{AI-based multi-module joint design in the mentioned methods\label{tab:table0}}
    \centering
    \begin{tabular}{c|c|c|c|c|c|c|c}
    \hline
        \textbf{Ref.} & \textbf{Pilot Design} & \textbf{CE} & \textbf{CSI Compression} & \textbf{Channel Coding} & \textbf{Channel Decoding} & \textbf{CSI Reconstruction} & \textbf{Precoding Design} \\ \hline
        \cite{CS_CsiNet} &   &  &  &  &   & \checkmark &   \\ \hline
        \cite{CFnet} &   &   &   &   &   & \checkmark &   \\ \hline
        \cite{PPPNet} &   &   &   &   &   & \checkmark &   \\ \hline
        \cite{CSINET_Plus} &   &   & \checkmark &   &   & \checkmark &   \\ \hline
        \cite{SwimTrans} &   &   & \checkmark &   &   & \checkmark &   \\ \hline
        \cite{ADJSCC} &   &   & \checkmark & \checkmark & \checkmark & \checkmark &   \\ \hline
        \cite{CEFNet} &   & \checkmark & \checkmark &   &   & \checkmark &   \\ \hline
        \cite{JFPNet} &   &   & \checkmark & \checkmark & \checkmark & \checkmark & \checkmark \\ \hline
        \cite{Yuwei_MU_Precoding} & \checkmark & \checkmark & \checkmark &   &   & \checkmark & \checkmark \\ \hline
    \end{tabular}
\end{table*}

In this article, we first introduce an AI-based one-sided CSI feedback method, wherein the DL model is only employed at the BS, and AI-based multi-module learning involving the CSI feedback. The different architectures discussed in this paper are summarized in Table~\ref{tab:table0}, with particular emphasis on the specific modules where AI techniques have been utilized as substitutes. In particular, architectures with multiple ticks cater for the joint design of multiple modules, which is what we introduce immediately afterward. For example, deep joint source-channel coding (DJSCC) can be exploited as the joint design of CSI compression and channel coding for the feedback channel. Lastly, we articulate the prospects and challenges pertaining to AI-driven CSI feedback tasks, taking into account both the AI limitations and the real-world implementation of AI models.

\section{AI-Based One-Sided CSI Feedback}
Given the broad challenges associated with AI models, it becomes essential to train and utilize different parameters for distinct CSI feedback scenarios, e.g., considering different wireless environments and compression ratios. However, the storage and deployment of numerous complex AI models at low-cost terminals, which are often limited in device memory and computational capabilities, pose significant challenges. 
To reduce the storage overhead, one approach is to store multiple models in a BS that has superior storage capacity. When the task at hand is performed, the BS would identify the current scenario and dispatch the suitable model to the UE. However, this method will unavoidably lead to additional spectrum resources' consumption. To enable a practical deployment and minimize the transmission overhead for the CSI feedback task, the AI model can be implemented solely on the BS, a scheme known as one-sided CSI feedback. Moreover, this one-sided CSI feedback architecture avoids joint training and additional collaboration overhead between the network and UE vendors, ensuring both the confidentiality of the model architecture and the privacy of the training data.

Some novel one-sided CSI feedback architectures were proposed in \cite{CS_CsiNet,CFnet,PPPNet}. The CS-CsiNet \cite{CS_CsiNet} implements CSI compression through linear projections and utilizes DL for CSI reconstruction. Using the codebook-based CSI feedback at the UE, the CFnet \cite{CFnet} adds a DL-based refine network at the BS side to enhance the accuracy of CSI reconstruction. This improvement is achieved by introducing environmental knowledge to refine the initial reconstructed CSI. However, these DL-based one-sided methods can
only compress the CSI matrix with a fixed compression ratio. The BS has to train and store several DL models to realize multi-rate CSI compression. CSI-PPPNet \cite{PPPNet}, i.e., a one-sided one-for-all framework for DL-based CSI feedback, allows a DL model to work with arbitrary CSI compression ratios. Specifically, the CSI is compressed simply via a small number of linear mapping at the UE, and is recovered at the BS in an iterative manner, which involves the use of a DL-based denoiser following the plug-and-play priors (PPP) framework. Notably, the training process remains independent of the compression process in CSI-PPPNet. As a result, this one-sided CSI feedback architecture can be applied across various compression rates, while the training and maintenance of the model can be handled only by the network vendor.


\begin{figure}[!t]
\centering
\includegraphics[width=3.4in]{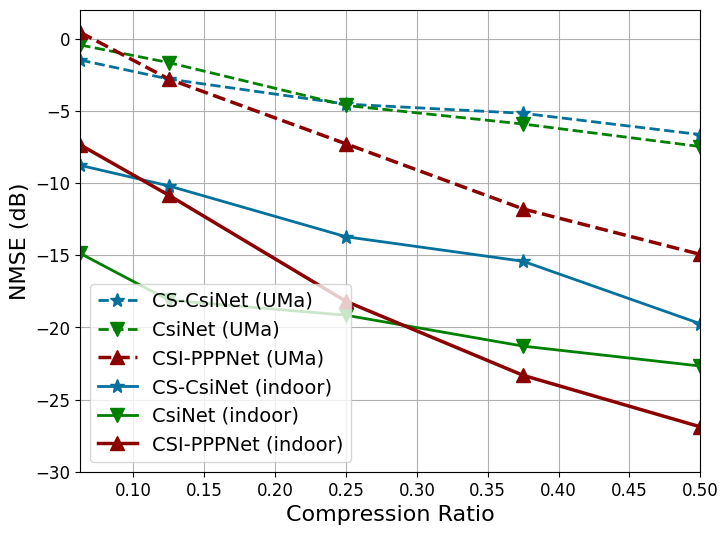}
\caption{Performance comparison of AI-based one-sided CSI feedback networks, i.e., CS-CsiNet and CSI-PPPNet, and AI-based two-sided CSI feedback network, i.e., CsiNet, for the indoor and UMa scenarios.}
\label{PPPNet}
\end{figure}

Figure~\ref{PPPNet} shows the normalized mean square error (NMSE) performance of AI-based one-sided CSI feedback networks, i.e., CS-CsiNet and CSI-PPPNet, and AI-based two-sided CSI feedback network, i.e., CsiNet, for indoor and urban macrocell (UMa) scenarios. CsiNet is a popular two-sided CSI feedback model used as a baseline algorithm \cite{CSINET_Plus, PPPNet, CS_CsiNet}. CSI-PPPNet achieves the lowest NMSE values for high compression ratios. Remarkably, with a compression ratio of 1/2, CSI-PPPNet can decrease the NMSE by over 5 dB, outperforming the other two methods in the UMa scenario. Note that this result does not suggest that one-sided models would always have better accuracy than two-sided models. The main advantages of the considered one-sided model are in the avoidance of collaboration between vendors and the model deployment. Although CsiNet shows good performance at compression ratios of 1/8 and 1/16 in indoor conditions, it requires a substantially larger number of model parameters in its two-sided model compared to CSI-PPPNet.


In the case of one-sided CSI feedback, the UE  has to merely preserve and return a single additional parameter to the BS to create the linear mappings. Furthermore, the suggested CSI-PPPNet possesses an attractive one-for-all property, i.e., only a single DL-based denoiser model is required for deployment for any compression ratios, substantially cutting down the quantity of models required for training and storage at the BS. To accommodate CSI feedback across varying compression ratios, the number of model parameters in CsiNet, CS-CsiNet, and CSI-PPPNet is 4095K, 2062K, and 175K, respectively. Specifically, the CsiNet requires 2033K parameters at the UE and 2062K parameters at the BS. The one-sided models CS-CsiNet and CSI-PPPNet both require one parameter at the UE, and 2062K and 175K parameters at the BS, respectively. The CSI-PPPNet only constitutes 4.3\% and 8.5\% of the parameter count in CsiNet and CS-CsiNet, respectively.

\section{Joint Multi-Module Learning With CSI Feedback}
CSI compression is only one module in the communication framework as shown in Fig.~\ref{scenario}. To exploit the full potential of MIMO technology and enhance the communication system throughput, various other modules in the communication system, including channel coding, CE, pilot design and precoding design, need to be jointly designed together with the CSI feedback task. In this section, we consider various joint multi-module learning along with the CSI feedback task.

\subsection{Joint CSI Compression and Channel Coding}
The CSI compression can be viewed as a source compression task, which can be designed independently of channel coding, following the paradigm of separate source-channel coding (SSCC).
Using AI-based source coding methods, the CSI is compressed by an encoder and reconstructed using the corresponding decoder. The configuration of the encoder and decoder can be flexibly adjusted, for instance, by leveraging convolutional neural networks (CNN) \cite{CSINET_Plus}, transformer networks \cite{SwimTrans}, and so on. Nonetheless, this configuration overlooks the influence of channel coding and the feedback channel on the reconstruction process. When the quality of the channel deteriorates to such an extent that it surpasses the processing capacity of the channel coding, the precision of the reconstructed CSI within the SSCC framework experiences a significant drop, a phenomenon referred to as the ``cliff effect''. The use of heavily distorted CSI for the creation of precoding vectors results in an undesirable reduction in the system throughput. While the hybrid automatic repeat request mechanism and other methods can alleviate this problem, they also cause a delay in the downlink CSI acquisition and an increase in the feedback overhead.

\begin{figure}[!t]
\centering
\includegraphics[width=3.45in]{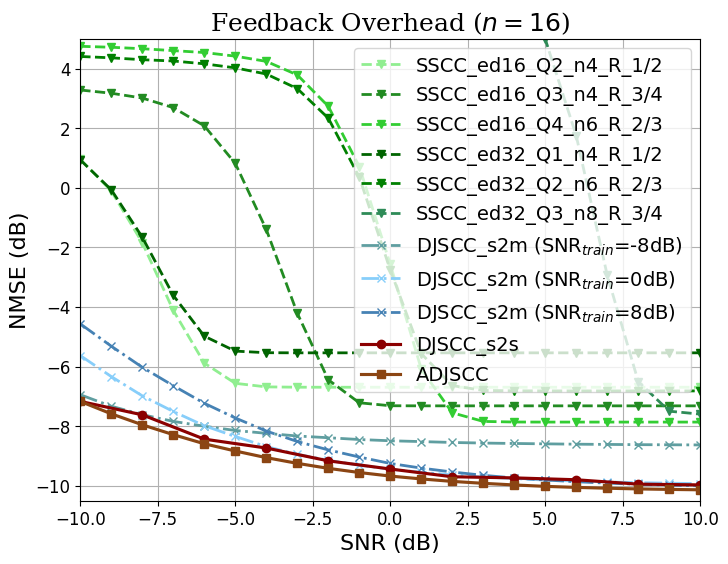}
\caption{Performance comparison between SSCC, DJSCC, and ADJSCC with the same feedback overhead. The label ``SSCC\_ed16\_Q2\_n4\_R\_1/2'' refers to the SSCC network where the encoder has 16 outputs of real number, 2 quantization bits, a QAM modulation order of 4, and a coding rate of $1/2$.}
\label{ADJSCC}
\end{figure}

To address the ``cliff effect'', a new CSI feedback architecture, coined as DJSCC, was introduced in \cite{ADJSCC}. This approach leverages DL to integrate source and channel coding, training the system with a dataset that encompasses both the source and the wireless channel. DJSCC incorporates the environmental characteristics of the uplink channel in its comprehensive training process, aiding in mitigating the ``cliff effect'' seen in SSCC. The performance of the SSCC and DJSCC networks for CSI feedback with the same overhead is shown in Fig.~\ref{ADJSCC}. Both networks are trained and tested on the same dataset, which is generated by QuaDriGa in a FDD indoor scenario. The number of feedback symbols $n$ is set to 16. The NMSE between the reconstructed CSI and the target CSI is utilized to evaluate the network performance. We follow the experimental settings and employ the network architecture in \cite{ADJSCC}. Using the Discrete Fourier Transform (DFT) matrix, the CSI in SSCC is first converted from the spatial-frequency domain to the delay-angle domain, which has a sparse pattern. After retaining only the essential information, the CSI is compressed using an encoder comprising multiple convolutional layers. The output of the encoder is fed back after quantization, channel coding, and modulation. The offset and the decoder module, both consisting of convolutional layers, are employed at the BS to offset the quantization loss and reconstruct the CSI, respectively. Notably, the output of the decoder is the truncated CSI in the delay-angle domain. To obtain full CSI in the spatial-frequency domain, a matrix of zeros must be appended after the truncated CSI for dimension recovery. 

To preserve as much useful data as possible in DJSCC, the DFT based domain transform in SSCC is replaced with an analytical transform network and a synthetic transform network \cite{ADJSCC}. Both networks are composed of convolutional networks to achieve upsampling and downsampling, respectively. The encoder and decoder for CSI compression are the same as those in the SSCC. The DJSCC network utilizes an end-to-end training method to update the network parameters while simultaneously taking into account the uplink channel environment knowledge and source information. In the training phase, SSCC trains the CSI compression and CSI reconstruction module without considering the uplink channel. In contrast, DJSCC trains the end-to-end network considering the uplink channel with a fixed uplink signal-to-noise ratio (SNR). For the testing phase, DJSCC is tested at the training SNR, referred to as ``single-point training, single-point testing (s2s)''. From the experimental results shown in Fig.~\ref{ADJSCC}, we can observe that the SSCC network suffers from a severe ``cliff effect''. DJSCC addresses this issue by yielding only a gradual rise in NMSE with the reduction of the SNR in the feedback channel, while enhancing the performance across all SNR levels.

On the other hand, the method of single-SNR training presents a problem for generalization, as indicated by the curves labeled "s2m" in the legend of Fig.~\ref{ADJSCC}. These curves are trained at a fixed $\rm{SNR}_{train}$ and then tested at various SNRs. It has been discovered that the DJSCC network performs optimally only when the testing SNR equals the training SNR. To address the generalization problem, an attention mechanism-based DJSCC network (ADJSCC) was proposed in \cite{ADJSCC}. An attention feature (AF) module was incorporated into the DJSCC network. With the AF module, ADJSCC can dynamically adjust the ratio of source coding and channel coding outputs based on the uplink SNR. Compared to the DJSCC network, the training SNR in ADJSCC is no longer fixed but randomly selected within a certain range. As illustrated in Fig.~\ref{ADJSCC}, incorporating the AF module allows the ADJSCC network to outperform various networks trained at different SNRs. 


\subsection{Joint CSI Compression, Channel Coding and Precoding}
As the number of antennas and subcarriers increases, achieving compressed feedback of full CSI with low feedback overhead remains a significant challenge. In view of the fact that CSI is exploited to design the precoding matrix in MIMO systems, it would be promising to jointly design the CSI compression, channel coding, and precoding module. In other words, we do not need to explicitly reconstruct the CSI at the BS but directly generate the precoding matrix according to the feedback information. This task-oriented multi-module end-to-end architecture could maximize the downlink sum-rate by transmitting not explicit full-channel CSI, but rather semantic information about the implicit CSI associated with the precoding. 

In a multiuser MIMO scenario, the precoding vector of one UE would affect its interference to/from other UEs. Hence, to maximize the downlink sum-rate, the precoding vectors are designed jointly after the BS aggregates the CSI of all UEs. The error in the reconstructed CSI affects the efficacy of the precoding design, and thus a precoding design module trained using perfect CSI does not result in optimal performance. To maximize the downlink sum rate (or equivalently the sum SE) some joint feedback and precoding networks (JFPNet) have been proposed to simultaneously optimize the feedback network and precoding design network \cite{JFPNet}. UEs are assumed to have perfect downlink CSI, which can be used to generate the eigenvalues and eigenvector matrix. Subsequently, the eigenvector matrix of each UE is compressed via the source-channel joint coding module, as in \cite{ADJSCC}. At the BS, the UEs' eigenvector matrices are reconstructed in parallel via the same decoder module as in \cite{ADJSCC}. The reconstructed eigenvectors are then used to design the precoding vectors. The precoding design module has two components, i.e., the direction design module and the power design module. The direction design module consists of multiple fully connected (FC) layers with a power normalization constraining its output, which is called JMP module. For the power design module, the perfect feedback eigenvalues and downlink noise power are used as auxiliary information, and an FC network is employed to allocate the power between different UEs. Finally, the overall precoding design is realized by multiplying the outputs from the aforementioned two modules.

\begin{figure}[!t]
\centering
\includegraphics[width=3.45in]{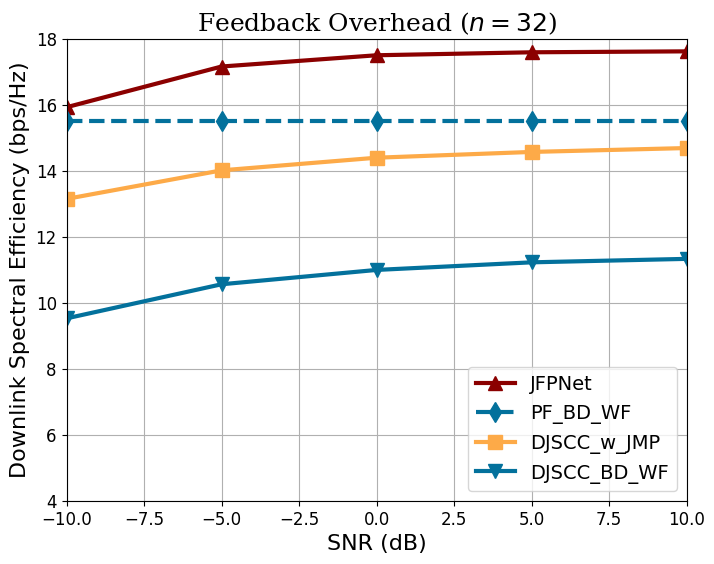}
\caption{Performance comparison: a joint training approach and a separate training approach are utilized for the DL-based CSI feedback module and precoding module, respectively.}
\label{JFPNet}
\end{figure}

Now, we show the gain for joint precoding design in the CSI feedback multi-module learning. The uplink and downlink channel datasets are created using QuaDriGa in the FDD UMa scenario. Initially, the downlink CSI undergoes preprocessing to generate the eigenvalues and the eigenvector matrix, and the DJSCC network is then used to compress the eigenvector matrix for feedback. The uplink channel SNR during the training phase spans a range from $\rm{-10}$ dB to $\rm{10}$ dB, following a uniform distribution and the feedback overhead is fixed to $n=32$ symbols. Figure~\ref{JFPNet} shows the downlink SE of different methods. The JFPNet, which jointly implements CSI compression, channel coding and precoding, achieves the highest SE at all SNRs.
The label ``DJSCC\_w\_JMP'' signifies that the JMP module for precoding design is trained separately with the DJSCC module for CSI feedback. Specifically, the DJSCC network is trained using the mean squared error (MSE) loss function for CSI feedback, while the JMP module is trained using  perfect CSI with the downlink SE as the objective. The label ``JFPNet'' indicates that the CSI feedback module and the precoding design module are jointly trained, with the training objective being the maximization of the downlink SE. Notably, ``JFPNet'' outperforms ``DJSCC\_w\_JMP'', suggesting that multi-module learning enables more effective extraction of semantic information related to the final task. The labels ``DJSCC\_BD\_WF'' and ``PF\_BD\_WF'' denote the DJSCC CSI feedback and perfect CSI feedback, respectively, both availing of the same traditional non-AI precoding design using the block diagonalization (BD) and the water-filling (WF) algorithms. Comparing ``DJSCC\_BD\_WF'' and ``PF\_BD\_WF'', a noteworthy observation is that the precoding design of traditional algorithms heavily relies on the reconstruction accuracy of CSI. With the same CSI feedback method DJSCC, the DL-based precoding design, i.e., ``DJSCC\_w\_JMP'', outperforms the traditional precoding design, i.e., ``DJSCC\_BD\_WF''. This result showcases that the the DL-based precoding is more robust to imperfect CSI. It is worth noting that the performance of ``JFPNet'' even surpasses that of the traditional precoding design with perfect CSI feedback. 

\subsection{Joint Channel Estimation and CSI Compression}
The CSI feedback methods introduced in the previous sections assume perfect CSI at the UE. In practice, pilot-assisted CE is widely used to obtain the downlink CSI, such that perfect downlink CSI at the UE cannot be guaranteed. The simple least squares (LS) algorithm usually exhibits limited CE accuracy, while CS-based CE offers improved performance by exploring the sparse channel structure. However, CS-based CE suffers from high computational complexity and the strict restriction on the sparse channel structure. The application of DL techniques introduces a new approach to solve the CE problem as a super-resolution problem or denoising problem \cite{pilot_CE}. In DL-based CE, a simple traditional CE algorithm is initially used to acquire a low-resolution channel matrix, such as the LS algorithm. Then, DL can be utilized to obtain a more accurate estimation of the channel matrix. 

Instead of designing the CE and CSI compression/reconstruction independently, CEFnet introduced a joint solution that realizes the estimation, compression, and reconstruction of downlink channels in FDD massive MIMO systems \cite{CEFNet}. To reduce the storage burden and computational overhead of the UEs, PFnet, a network that does not require explicit CE was also proposed in \cite{CEFNet}. In PFnet, CE is no longer applied at the UE. Instead, the received pilots is directly compressed. The powerful learning and nonlinear mapping capabilities of DL enable the completion of the CSI feedback task without the acquisition of the entire downlink CSI at the UE side. For both CEFnet and PFnet, the CSI reconstruction accuracy is superior to that of traditional algorithms. Compared to CEFnet, although the performance of PFnet is slightly degraded, the PFnet model contains fewer parameters and requires less storage space, making it more suitable for low-cost UEs. Pilot design for CE and precoding design were further considered in the joint multi-module learning framework in \cite{Yuwei_MU_Precoding}, where implicit CSI feedback was utilized. The experimental results in \cite{Yuwei_MU_Precoding} demonstrated an enhanced downlink throughput for the joint multi-module learning framework including pilot design, CE, CSI feedback, and precoding.


\section{Opportunities and Challenges}
DL-based CSI feedback in FDD massive MIMO systems offers reduced overhead and computational complexity. Employing a one-sided model and joint multi-module learning can facilitate the practical deployment and boost the end-to-end performance, respectively. Nevertheless, several challenges persist, both from the standpoint of AI technology and the real-world implementation of AI models for CSI feedback tasks. In this section, we will elaborate on these challenges.

\subsection{Inherent Limitations of AI}
Although AI methods offer many advantages, the effectiveness of an AI model greatly depends on the quality of the dataset. For the joint CSI compression and channel coding task, the datasets consist of downlink CSI and uplink CSI, which serve as the source to be compressed and the feedback channel, respectively. For joint multi-module training, the downlink CSI will be used as the channel for the downlink pilot transmission and for evaluating the downlink precoding performance. Existing datasets are primarily statistical channels generated by simulators, assuming that channels of different cells are independent and identically distributed. However, this assumption deviates from practical conditions and fails to depict the different distribution of channel information caused by different actual environments. Mismatched datasets will cause the trained model to fail in achieving the expected performance, thereby degrading the system throughput. One solution is to use real measurement data. However, the large amount of data required for model training would entail a huge cost to future measurement trials. Thus, exploring methods to maximize the use of restricted data for data augmentation or data mining in model training represents a key area for future research. Another solution is to first train the AI model offline using simulation data and then fine-tune the network parameters online in a specific environment using small datasets of real measurements.

Both approaches struggle to address every communication scenario comprehensively, and developing distinct models for each scenario results in prohibitive cost in terms of training and storing the models. Thus, scalable model generalization remains an important area for future research. For example, in \cite{ADJSCC}, the SNR information was utilized as auxiliary information to instruct the network through an attention mechanism. This method enables the network to adapt autonomously to different SNRs, allowing for the maintenance of a single network across various SNR levels in a specific scenario.

\subsection{Challenges in Practical Deployment}

In 6G wireless networks, expectations for mobile broadband services are elevated, requiring real-time and precise CSI. However, the channel changes with variations in the scattering environment and the location of the UEs. Delays in CSI feedback scheduling mean that the CSI becomes outdated, particularly in scenarios of high mobility, exacerbating the problem of channel aging.

To overcome the channel aging effects, channel prediction can be employed, which utilizes the historical channels to predict channels in the next several frames. Recently, AI-based channel prediction has been explored in some studies and achieved satisfactory performance. Specifically, in \cite{CSI_pre}, the self-attention mechanism was employed to leverage the temporal correlation of the historical CSI to forecast future CSI. Temporal domain CSI compression and prediction for enhancement using the UE-sided model has also been studied in the 3GPP Release 18 and Release 19. The joint CSI compression and prediction framework has emerged as a key research direction to alleviate the performance loss caused by channel aging. 3GPP has been studying temporal domain CSI compression using the two-sided model, to alleviate the performance loss caused by channel aging. For joint CSI compression and prediction, assessing the benefits of various model deployment scenarios and network configurations is crucial to enhance the accuracy of AI-driven CSI reconstruction in dynamic channels. Based on preliminary results in the 3GPP study, it can achieve promising gain comparing with using outdated CSI.

In real-world applications of AI models for CSI feedback tasks, due to the restricted generalization ability of these models, maintaining service quality requires investigations into live model monitoring and timely model switching. The design of appropriate quality indicators for model monitoring requires more investigation. When employing final key performance indicators (KPIs), e.g., system throughput, as monitoring metrics, it is challenging to precisely identify the problems across multiple modules. When employing intermediate KPIs, like CSI reconstruction accuracy for monitoring, the placement of model monitoring tasks must be carefully evaluated. If deployed at the UE, proxy models for CE or the compression of the original CSI become necessary, albeit at the cost of increased hardware complexity or additional transmission overhead. An alternative approach to model monitoring involves evaluating if the current feedback CSI substantially diverges from the training dataset, which indicates the necessity for model switching. This technique can support a reduction in the transmission overhead and an efficient model monitoring. 

\section{Conclusion}
In this article, we investigated the recent developments of DL-based CSI feedback from the perspectives of one-sided model and joint multi-module learning. Firstly, we introduced a one-sided CSI feedback architecture that only replaces the traditional decoding module with a DL module. Subsequently, we delved into CSI feedback architectures, jointly learned with various modules, e.g., channel coding, precoding and CE. For example, the joint design of the CSI feedback and channel coding via DJSCC overcomes the ``cliff effect'' observed in the traditional SSCC-based CSI feedback methods. 
Unlike the traditional approach of designing each module independently, multi-module joint design leverages DL techniques to jointly optimize the various modules and exploit the interdependencies between different modules, and can consequently enhance the overall system performance and training efficiency. 
\bibliographystyle{IEEEtran}
\bibliography{myref}

\end{document}